\def\year{2020}\relax
\documentclass[letterpaper]{article} 
\usepackage{aaai20}  
\usepackage{times}  
\usepackage{helvet} 
\usepackage{courier}  
\usepackage[hyphens]{url}  
\usepackage{graphicx} 
\usepackage{graphicx}  
\newcommand{\citet}[1]{\citeauthor{#1}\shortcite{#1}}
\newcommand{\citep}{\cite}
\newcommand{\citealp}[1]{\citeauthor{#1} \citeyear{#1}}

\title{Combating The Machine Ethics Crisis: An Educational Approach}
\author{Tai Vu\\ 
Department of Computer Science\\ 
Stanford University\\
taivu@stanford.edu 
}
 
\begin{document}

\maketitle

\begin{abstract}
In recent years, the availability of massive data sets and improved computing power have driven the advent of cutting-edge machine learning algorithms. However, this trend has triggered growing concerns associated with its ethical issues. In response to such a phenomenon, this study proposes a feasible solution that combines ethics and computer science materials in artificial intelligent classrooms. In addition, the paper presents several arguments and evidence in favor of the necessity and effectiveness of this integrated approach. 
$\newline$
\end{abstract}

\section{The Age of Intelligent Machines} 
$\newline$
Machine learning, a primary branch of Artificial Intelligence (AI), is regarded as one defining technological frontier shaping the 21st century. The advent of intelligent machines that learn and evolve through experience has pervaded a variety of sectors, including science, economics, finance, laws, transportation, and medicine. \\

In accordance with this trend, tech giants like Google, Facebook, Microsoft, and Amazon strive to take the lead in the machine learning race. According to McKinsey Global Institute’s report, investment in AI-aided projects in 2017 was between $\$18$ billion and $\$27$ billion \citep{mckinsey:2017}. Cutting-edge machine learning programs such as spam filtering, image recognition, recommender systems, and text translation are becoming dominant in a whole host of business industries. \\

However, as machine learning advances, it has aroused growing concerns over its ethical and societal implications, ranging from unfairness to transparency. To address these existing problems, some renowned universities have incorporated ethics contents into their AI curricula. However, more academic institutions across the world need to adopt this ethics training approach to equip machine learning developers with the knowledge needed for handling ethical issues that arise from their AI-enabled products. \\

\section{Machine Learning and Its Beneficial Aspects}
$\newline$
Before delving into the discussions of machine ethics, it is essential to comprehend machine learning’s mechanics. In simple terms, machine learning is a state-of-the-art technology that extracts patterns and trends from datasets. By analyzing an array of observations, the systems employ analytical, probabilistic, and statistical models to devise an estimated mathematical function mapping the attributes of these samples to the needed output. Afterwards, the machine uses this correlation to make predictions on new observations. For instance, a medical diagnosis algorithm trained on gigantic records of patients can forecast the likelihood of individuals having cancer cells based on their age, occupation, daily activities, eating habits, and their families’ medical history. \\

The key advantage of machine learning over conventional computer programs is its “learning” nature. Traditional algorithms rely on sets of hand-coded instructions to execute tasks step by step. In contrast, machine learning engineers remove these task-specific procedures and instead show the systems large collections of data points as problem-solving examples. Subsequently, these programs can teach themselves to infer hidden correlations amongst the data and predict outcomes on unseen inputs. \\

Machine learning’s predictive power plays a pivotal role in business and economic settings. In fact, governments leverage machine learning to foresee economic downturns and then enact fiscal and monetary policies accordingly. In addition, many companies employ these intelligent programs to analyze their customer data, gaining invaluable insights into customer preferences and market demands. This information supports managers to make informed sales and marketing decisions under time constraints. \\

\section{Ethical Consideration: \\ What Could Go Wrong?}
$\newline$
Despite the aforementioned promises, over-reliance on machine learning comes with a number of ethical problems. These existing limitations demonstrate an urgent need for practical approaches and solutions. \\

One pressing issue regarding machine-driven applications is the cost of fairness. In fact, many intelligent systems exhibit signs of unfairness and bias against certain human groups. This phenomenon is illustrated by the case of Amazon’s autonomous recruitment tool, which favored male candidates over their female counterparts. Specifically, this system “reportedly downgraded resumes containing the words "women's" and filtered out candidates who had attended two women-only colleges” (Hamilton, 2018). In addition, racial discrimination occurred with a risk-assessment product called COMPAS, which assisted judges in Broward County, Florida to select prisoners to let out on bail. A 2016 report by The ProPublica highlighted “significant racial disparities” against black people: “The formula was particularly likely to falsely flag black defendants as future criminals, wrongly labeling them this way at almost twice the rate as white defendants” \citep{cohen:nd}. \\

Furthermore, machine-driven data analytics is apt to violate citizens’ privacy and liberty. For example, Facebook’s photo-tagging system is called into question, since it identifies the identity of individuals in any images on Facebook, collecting them to build massive datasets of photos. The facial recognition program draws privacy concerns, as Facebook can track its users’ daily activities via their pictures. Additionally, this photo-scanning feature may be exploited for ungraceful purposes like harassment or bullying. A harasser can simply upload an image of their target, and Facebook “will recognize their face and ping that user, doing the harasser’s work for them” \citep{vincent:2017}. \\

Hence, an ethical lens urges caution against the unintended moral consequences of applied machine learning such as unfairness and the violation of privacy. It is necessary to examine how these challenges disrupt people’s daily lives and how to effectively handle such ethical repercussions. The search for viable solutions to these questions demands a collaboration of experts from a wide range of branches, including educators and academic institutions. \\

\section{Reforming AI Education with Ethics Training}
$\newline$
It is commonly believed that legislative regulation is the most effective remedy to resolve the machine ethics crisis. Elon Musk, co-founder and CEO of SpaceX, Tesla and OpenAI, urged US governors to quickly regulate AI software before it threatens human civilization \citep{condliffe:2017}. From this angle, governments should propound moral codes for assessing machine learning programs and outlaw the use of systems that do not conform to these standards. It is true that this approach may partially eradicate unethical machine-driven tools, thereby protecting users’ benefits and human rights. \\

Nonetheless, this proposal seems rather impractical, given the technological complexity and abstraction of machine learning. It is observed that the majority of policy-makers lack technical backgrounds in computer science in general and artificial intelligence in particular. US Congress’ questions to Facebook’s CEO Mark Zuckerberg regarding the firm’s data management suggested a lack of fundamental knowledge of modern technologies and straightforward business models derived from them \citep{stewart:2018}. Meanwhile, machine learning is an advanced field that requires sophisticated understandings of mathematics, statistics and computer science. These fields take adept software developers years to master. There are more than thirty distinct machine learning models, each of which possesses different natures and requires different policy approaches. Without essential expertise, governors face enormous difficulty devising legal frameworks for the use of intelligent machines. These regulations may not be applicable to real-world AI-enabled products. What is more, artificial intelligence is developing at a pace that policy-makers can hardly keep up with. Over the past five years, machine learning has progressed by leaps and bounds, beating humans in multiple spheres, including language translation, image recognition, and cancer detection, and this trend is projected to continue at an accelerating rate \citep{shoham:2018}. In contrast, the Congress often needs a long time to discuss, evaluate and enact a legislative proposal. Technology would escalate to a new level of sophistication and advance, making the policy obsolete. \\

On top of that, while the enforcement of these regulations filters out improper algorithms, it can limit machine learning applications and hinder the power of artificial intelligence. The strict control of data use, for example, can result in a scarcity of massive datasets for the training and validation of intelligent machines. The process of testing and assessing AI programs in government agencies is also highly time-consuming. According to a 2016 government report on AI policy, excessive or inappropriate regulatory responses can create “bottlenecks” that slow down the adoption and development of AI innovations \citep{president:2016}. Furthermore, the policy approach is insufficient, as it fails to address the root of the problem. Eventually, it is machine learning engineers who design, implement and execute the automated programs. Therefore, it is ideal to equip AI developers with proper ethical foundations, which allow them to produce highly accurate machine learning software that complies with ethical and social standards. \\

With growing preferences for AI-driven systems comes great demand for competent machine learning engineers, which further calls for ethics training in current AI-focused curricula. An employment report by Indeed reveals that, “with an average salary base of $\$146,085$ and a whopping $344\%$ growth in job postings, machine learning engineer is an extremely promising position” \citep{indeed:2019}. These career prospects accompany a surge in machine learning education: “2017 introductory AI enrollment was $3.6\times$ that of 2012, while 2017 introductory ML course enrollment was $5\times$ that of 2012” \citep{shoham:2018}. However, the ethical components of these technical classes have been underappreciated. A survey on machine learning engineers shows that roughly $12\%$ of them regard ethics as important, while only $5\%$ employ ethical knowledge when pursuing their AI careers \citep{wollowski:2016}. If this phenomenon remains unaddressed, a myriad of machine-aided products will be created without moral consideration, thereby exacerbating the current situation. Thus, these statistics illustrate the role of ethics contents in universities’ AI curricula, as such a practice would transform the minds of millions of machine learning developers who drive future artificial intelligence advances. \\

\section{Promoting Transparency}
$\newline$
A growing ethical worry amongst governments and citizens is a lack of algorithmic transparency in AI-enabled systems. An ethics-oriented AI curriculum would provide machine learning engineers with an ethical mindset, which encourages them to make their programs transparent to the public’s eyes and resolve potential moral concerns. \\

The need for explainable machine learning is worthy of consideration, since opacity is a common problem amongst many learning models. Since a multitude of currently-used algorithms are derived from complex statistical and mathematical principles, it is tough for their developers to explain their mechanics to the general public. The notion of such “black box” algorithms exceptionally holds for deep learning, an upgraded version of machine learning. This model is implemented on neural networks, which encompass thousands of interconnected neurons distributed into a chain of sequential layers. This structure enables deep learning to incorporate many machine learning algorithms into a single framework, handle massive datasets, and uplevel its predictive accuracy. However, this productivity comes at the cost of transparency. “Once [the network] becomes very large, and it has thousands of units per layer and maybe hundreds of layers, then it becomes quite un-understandable,” suggests Jaakkola, an MIT computer science professor \citep{knight:2017}. In other words, deep learning is a “dark black box”, even for capable software engineers and computer scientists. Recognizing such complexities, many AI practitioners currently employ available open-source deep learning libraries, such as Keras, Pytorch, TensorFlow, Scikit-learn, MXNet, and Caffe, without paying attention to their inner workings \citep{lorica:2019}. What they frequently do is to plug in inputs, wait for their computers to run and then get desired outputs. \\

The “black box” nature of machine learning algorithms has several detrimental effects. Although they seem to perform competently in the meantime, there is no guarantee that an unethical machine-driven decision would not occur in the future. When that happens, AI engineers will have a tough time understanding its underlying causes and alleviating its consequences. Additionally, AI’s opaque property draws moral criticisms from governors and the public. As an illustration, job candidates who get rejected by machine-based recruitment systems and customers who are denied loans by banks’ automated profiling processes may question whether these programs treat them unfairly and contain biases. In response to such matters, the European Union introduced the General Data Protection Regulation, censoring the use of inscrutable artificial intelligence techniques. A paper by Bryce Goodman and Seth Flaxman, two researchers at the University of Oxford, illustrates that the “right to explanation” in this new framework, which allows users to ask for explicit explanations of algorithmic decisions, poses a challenge to many “black box” algorithms like random forests, support vector machines, and neural networks \citep{goodman:2017}. Even though deep learning offers high accuracy and efficiency, it cannot be put into practice. \\

The case of opaque machine learning programs drives the necessity of educated AI developers who has the ability to explain the functioning behind training algorithms. That is one learning goal of “CS181: Computers, Ethics, and Public Policy”, a popular course at Stanford University. This type of computer science classes raises students’ awareness of ethical issues stemming from non-transparent machine learning \citep{cs181:2020}. Such a practice motivates young computer scientists to put efforts into understanding factors that constitute the interpretability of a model and aiming for the explanation of these automated processes. With these toolkits, they can design and utilize complicated architectures like deep learning with great confidence, gain users’ trust, and lessen public ethical worries. \\

A large number of machine learning practitioners can reap certain benefits from the widespread teaching of these algorithmic transparency contents. Specifically, they are equipped with practical ways of purposefully developing and deploying interpretable models. Engineers may limit the number of variables and parameters in the training phase, which diminishes the number of uncovered correlations. Additionally, they can opt for applied models with comprehensible learning outputs like decisions tree, which allows users to easily walk through branches of a tree structure without getting confused by obscure parameters \citep{selbst:2018}. In addition, the ethics knowledge acts as an incentive for AI developers to delve into the internal process of “black box” architectures like deep neural networks rather than ignoring them and focusing entirely on inputs and outputs. By acquiring in-depth understandings of the models, they are able to simplify them and study practical tools to deliver the outcomes of these programs to unfamiliar stakeholders. As the opacity is alleviated, engineers also face fewer hurdles searching for and fixing problems like unfair selections and biases, thereby mitigating existing moral concerns over these systems. \\

What is more, the teaching of machine-related transparency benefits policy-making practices. By demystifying how intelligent machines reach final answers, developers can aid governmental agencies to make policy-related choices on the validity and applicability of training models and assess their ethical impacts on society. In this way, high-performance programs like deep learning can be permitted, while their functionality is kept under the government’s control. One may propound an alternative idea of instructing policy-makers the mechanics of machine learning, so that they can make decisions without AI engineers’ support. However, this measure is infeasible, since governors with no technical background would struggle with extensive mathematics, computer science and statistics know-how. The proposal is also unnecessary, since policy-makers only need to comprehend the rationale behind machine-driven decisions instead of in-depth mathematical formulas and concepts. Hence, it seems practical to introduce the knowledge of transparency to software developers. Having undertaken years of intensive training, they have fewer obstacles when digging deep into the functioning of machine learning models and translating it into simple words. \\

\section{Heightening Developers’ Data Integrity}
$\newline$
In addition to resolving the question of algorithmic transparency, teaching ethics in AI classes can stimulate data integrity amongst machine learning developers. This improvement plays an indispensable role in combating ethical problems of intelligent machines, because training data is a principal part of the learning process. \\

One critical pattern that AI practitioners need to be aware of is unintended biases in the datasets. In fact, majority groups often dominate the datasets, and this dominance can be amplified in the training process, which contributes to highly biased outcomes. One such case is indicated in the research “Gender Shades” by MIT Media Lab. By evaluating facial recognition products by IBM, Microsoft, and Face++, they pointed out key discriminations in terms of gender, skin type, and ethnicity. In particular, “$95.9\%$ of the faces misgendered by Face++ were those of female subjects,” while “$93.6\%$ of faces misgendered by Microsoft were those of darker subjects.” Such flaws arise as the training data is “overwhelmingly composed of lighter-skinned subjects.” Those “substantial disparities” in the computer vision systems demand the “urgent attention” of AI developers to the importance of cleaning hidden biases in the input data \citep{buolamwini:2018}. \\

To handle such issues, machine learning instructors should teach their students a code of ethics and fairness in processing data. This approach helps learners accumulate hands-on understandings of the primary sources of biases in training datasets. Specifically, the data may lack information about minority groups like African-American residents or contain prejudices coming from the history and culture of human beings. Without learning these concepts, engineers would be likely to neglect potential biases and plug the raw datasets directly into AI-enabled systems in the face of the pressure of quickly extracting useful insights from their industries. Consequently, their algorithms would inherit the discriminations and produce flawed outcomes, which can scale uncontrollably over computer systems. By contrast, research has shown that possessing ethical knowledge allows developers to withstand pressure from business concerns and appreciate the significance of cleaning data and removing biases \citep{dodig:2003}. \\

Several skeptics may believe that the awareness of data biases is of no help, as these characteristics and preferences are the nature of real-world data and are challenging to address. However, these critics overlook the fact that students in ethics-oriented AI courses grasp typical types of discrimination, including gender, race, skin type, and ethnicity. In many cases, they can calculate simple statistics to target these kinds of biases. Besides that, the teaching of ethics content sharpens machine learning engineers’ competence in high-level data preparation and bias detection. A fruitful model is “CS294: Fairness in Machine Learning” at the University of California Berkeley. This class facilitates their students’ discussion of data and unfairness in social contexts and enables them to harness advanced statistical modeling and causal inference methods, such as Simpson’s paradox, measurement theory, sampling, and unsupervised learning, to attack the core of these flaws \citep{cs294:2017}. If such a format is popularized, all AI students will develop a penchant for handling the data preparation process and uncovering biases properly. This way, machine neutrality is guaranteed and a major element of machine ethics crisis is mitigated. \\

\section{Ethical Designs of Machine Learning}
$\newline$
Asides from hidden data biases, ethical training in AI courses can guide ethical decisions throughout the development stages of computer programs, so that machine learning engineers have the ability to make moral algorithmic designs and implementation choices. \\

Ethical dilemma facing developers’ programming process is an important subject that AI courses in some US universities start to cover. For instance, the class “The Ethics and Governance of Artificial Intelligence” at Massachusetts Institute of Technology allows students to scrutinize “The Moral Machine experiment”, an online platform collecting public viewpoints on how self-driving cars should distribute the harms amongst distinct stakeholders in car crashes \citep{mit:2017}. Such techniques grant students access to global preferences including sparing more lives and sparing young lives, which are “essential building blocks” when they construct their algorithms. It is also important for them to obtain key variations by gender, religiosity, geographic features, and economic strengths: “both male and female respondents indicated preference for sparing females, but the latter group showed a stronger preference ... the preference to spare younger characters rather than older characters is much less pronounced for countries in the Eastern cluster, and much higher for countries in the Southern cluster” \citep{awad:2018}. \\

Some critics may argue that knowledge of diverging ethical perspectives is unnecessary for machine learning engineers, since their job is solely to devise high-performance models. This stance is superficial, because without understandings of these diverse standpoints, developers are inclined to make unreasonable coding assumptions that match their own sets of moral values, and the final outputs may contrast with others’ social norms and rules. They may also follow traditional optimization purposes, and even though their products like driverless cars maximize the number of protected people, they do not take into account varied ethical standards and thus trigger ethical concerns in different regions. Therefore, AI courses should instruct students to grasp conflicting viewpoints about ethical codes for machine learning, which is attributable to marked discrepancies in genders, cultural values, and living standards. Research in ethics training demonstrates that this expertise “[enables] the AI [practitioners] to reach a more ethically comprehensive position, allowing [them] to deploy familiar modes of reasoning while challenging [them] to look beyond [their] own utility and personal concerns" \citep{goldsmith:2017}. \\

Furthermore, AI professors ought to provide their students with ethical frameworks to reason about their algorithmic designs. This way, students are motivated to delve into machine ethics, looking at various schools of ethical thought from philosophical, historical, cultural, economic and legal standpoints. Therefore, young machine learning engineers are able to develop a comprehensive roadmap of ethical theory and determine ethical choices introduced in their machine learning projects. In fact, the effectiveness of in-depth moral reasoning is substantiated in a paper on ethics education: “By understanding the reasoning structure of the different theories, a practitioner is better equipped to follow the ramifications of her own values and judgments, and - once she has seen their implications - to reconsider those judgments and values” \citep{goldsmith:2017}. For that reason, Stanford University offers the interdisciplinary course “CS122: Artificial Intelligence - Philosophy, Ethics, and Impact” that examines ethical analysis in relation to autonomous machines. This is a striking example to follow, since it arms students with intellectual tools and ethical foundation to think critically and analytically about their implementation decisions and “successfully navigate the coming age of intelligent machines” \citep{cs122:2014}. \\

\section{The Merits of an Integrated Approach}
$\newline$
While the introduction of ethics in AI classrooms is needed, the way of delivering such knowledge to machine learning students is equally important. Some educators may believe that universities should force AI students to take separate ethics courses as part of their major requirements instead of mixing computer science and ethics. These specifically designed courses allow learners to interact with prominent, knowledgeable experts and develop specialized understandings of ethical reasoning from the ground up. On the other hand, these special courses seem conceptually demanding and less appealing to science-oriented students who want to get a taste of how ethics is correlated to machine learning products. A case study has illustrated that this approach may be counterproductive. As computer science students often experience intense workload, “even the most conscientious students would begin skipping classes and not studying if they thought they could get away with it” \citep{unger:2005}. \\

In contrast, blending ethics into machine learning classes is a more captivating and effective measure. One clear benefit is that when AI professors teach computer ethics, they serve as role models who employ moral reasoning to enhance their machine learning algorithms. Additionally, the course staff can invite renowned ethics faculty members to instruct in-depth ethical theories, thereby guaranteeing the quality and depth of moral contents for their students. Meanwhile, since these classes do not emphasize the pure modes of ethical thought and the evolution of human morals, learners are not obligated to read wordy textbooks and monotonous papers on history, humanities, and philosophy. Instead, students’ work focus on real-world ethical and technological implications on different parties, including their jobs and lives, so this emphasis cultivates their interest in learning the subject. \\

On top of that, these integrated classes can pick a discussion-based format, which facilitates students’ collaborative learning experience. For example, in “CS181: Computers, Ethics, and Public Policy” at Stanford University, instructors often pose engaging questions and provide guidance for learners to formulate answers in groups. Students can play the roles of distinct stakeholders like computer scientists, policy-makers, businesses, and ordinary citizens and discuss with their peers about real case studies in ethical codes of machine learning forecasts \citep{cs181:2020}. Empirical research has indicated that these instructor-guided discussions and role-playing activities in ethics classrooms act as catalysts to “productive conversations”, “cogent arguments”, and the acquisition of diverse standpoints, which positively correlate to students’ efforts and academic performances \citep{quinn:2006}. \\

What is more, homework assignments should combine technical problem-solving with ethical reasoning to optimize learners’ scholastic competence. For illustration, CS181 offers coding challenges that require students to incorporate algorithmic fairness into their AI-aided software \citep{cs181:2020}. For computer science majors, these practical problems are more engaging than reading and analyzing lengthy research papers, while still meet the goal of ethical training. Besides that, students gain a fresh outlook on designing actual machine learning algorithms under ethical consideration. A case study has suggested the advantage of combining coding skills with ethical analysis, which supports students to “connect their roles as professionals with their roles as moral agents” \citep{alenskis:1997}. Thus, this is a promising model for other colleges to follow. \\

Another strategy to boost the effectiveness of machine ethics education is called Embedded EthiCS, currently experimented at Harvard University. This program embeds ethics mini-modules into AI courses, giving students practical competence in thinking through ethical challenges \citep{grosz:2018}. One merit of this approach is that students’ grasp of new machine learning algorithms is accompanied by their ethical implications, which give learners comprehensive insights into the use cases of the technology. In addition, by partitioning ethical thoughts into small components in different courses, this curriculum design removes the barriers of theoretical ethics for computer science undergrads, because it allows for gradual, slow-paced acquisition of ethics contents, as suggested by education research \citep{alenskis:1997}. This practice ultimately increases the amount of ethical analysis gained in the long run as students fulfill their AI specialization. Moreover, this strategy brings ethics to a wider range of machine learning students, since computer science courses often have large capacities, compared to limited spots in small ethics-focused classrooms. \\

\section{Guaranteeing Social Responsibility}
$\newline$
In addition to dealing with ethical issues, the inclusion of ethical topics in AI courses helps evoke social responsibility amongst young researchers. With machine learning skills and an awareness of moral and social challenges facing humanity, developers are motivated to put their knowledge into practice and take the initiative to better their community. To fuel that inspiration, Stanford University offers “CS21SI: AI for Social Good”, which dives into fundamental machine learning techniques and their ethical and societal implications \citep{cs21si:2019}. Panel discussions with guest speakers from academia and industry broaden students’ understanding of the peril of biased and immoral machine learning and the merits of using AI in a socially conscious manner. Afterwards, students work on hands-on programming projects addressing severe ethical and social issues in various branches, not limited to the tech sector. Not only does this approach challenge students to think about their roles, missions and impacts on society, but it also inspires them to apply their machine learning expertise in social good domains, including education, government, and healthcare. Thus, this class is an effective model that every AI curriculum should adopt. \\

Outside the traditional classroom setting, academic institutions should promote socially conscious mindsets by supporting student-run groups focused on moral computer science in general and ethical AI in particular. Research has suggested that student organizations substantially contribute to students’ involvement in undergraduate education and reinforce their academic knowledge \citep{nadler:1997}. To that end, Stanford has invested in several clubs like CS + Social Good and EthiCS (\citealp{cssg:nd}, \citealp{wagner:2017}). These groups create open conversations where their members voice their viewpoints about core ethical issues of intelligent machines. CS + Social Good also partners with research labs and companies, enabling students to conduct real-world ethical AI projects in social good spaces. Such clubs complement formal AI and ethics education, since they sharpen students’ ethical reasoning skills and encourage students' ethical behaviors and responsibilities of returning back to their communities. \\

\section{The Way Forward}
$\newline$
In summary, incorporating ethics into machine learning curricula is of paramount importance to addressing current AI-related ethical issues. This model proves fruitful in some top-notch universities, and it needs to be widely spread to other academic institutions. By fostering ethical education in the AI sector, universities and colleges can produce socially minded machine learning engineers, who can drive ethically sound machine-aided products and mitigate moral concerns. In the long run, they will shape technological innovations, open the door to new data science revolutions, and bolster the progress of artificial intelligence and human intelligence.

\newpage

\bibliographystyle{aaai}
\bibliography{pwr1}

\end{document}